\documentclass[3p]{elsarticle}
\usepackage{epsf,amsmath,graphicx,subcaption,amssymb,url,hyperref}

\begin{document}
\begin{frontmatter}
\title{Limits on the Weyl meson parameters due to Fermi-LAT gamma-ray observations}
\author{ Gopal Kashyap \fnref{myfootnote}}
\address{Department of Physics, IIT Kanpur, Kanpur 208 016, India}
\fntext[myfootnote]{email:gopal@iitk.ac.in}
\begin{abstract}
We use gamma-ray observations by the Fermi Large Area Telescope (Fermi-LAT) to impose limits on the properties of the 
hypothetical particle Weyl meson. Such mesons arise in theories which display local scale invariance and act as dark matter candidate.
In a generalized locally scale invariant Standard Model it can decay or annihilate to the photon pair through the Higgs channel. 
We find that Fermi-LAT observations severely constrain the parameters of the Weyl meson.
\end{abstract}
\end{frontmatter}
\section{Introduction}
In recent years the hunt for the existence of dark matter (DM) in the Universe has increased significantly. 
There are many groups looking for the signature of dark matter existence in the Universe either 
directly or indirectly\cite{cogent,cdms,cresst,dama,pamela,ams,bringmann}. There also exist many theoretical models for the dark matter 
particles\cite{dm,jungman,bergstrom,feng,Bertone1}. The Weyl meson, which is a gauge particle of the local scale 
symmetry introduced by Weyl\cite{weyl} in 1929, 
is also considered as a potential Dark matter particle. The local scale invariant theories have many cosmological
and particle physics applications and are well studied in the literature \cite{deser,sen,dirac,utiyama,Freund1,hayashi,hayashi1,paddy,ranganathan,
rajpoot1,nishino,rajpoot3,quiros,bars,bars1}. The mechanism for breaking this symmetry has been
 discussed in the Ref.\cite{jain1,pseudoscale,brokenscale}. However, in the model containing only one Higgs doublet, the Higgs 
particle gets eliminated from the particle spectrum \cite{cheng,hcheng}. As now the Higgs particle has been found in the Large Hadron Collider (LHC) 
and is also required to make the theory perturbatively reliable\cite{joglekar,cornwall}, the local scale invariant theories need to be 
modified to incorporate the Higgs particle. A generalized locally scale invariant theory including extra one scalar field 
has been discussed in the Ref.\cite{quantumweyl,padilla}. 
In the model discussed in Ref.\cite{quantumweyl}, the Higgs particle remains in the physical particle spectrum and
also the Weyl meson gets weakly coupled to the Higgs boson. The Weyl-Higgs coupling may leads to many cosmological and collider
implications, as shown in Ref.\cite{GK}.

Till now there are no conclusive observations to infer the dark matter presence in the Universe. As it does not 
interact, or may interact very weakly, with SM particles, it is very difficult to find any direct evidence of the dark matter particles.
 The possible indirect evidence of dark matter presence in the Universe may be 
the observation of annihilation or decay product of these particles. It was shown in Refs.\cite{snellman} that the 
annihilation of dark matter particles into photon pair can produce the monochromatic gamma rays. Hence a clear peak, 
in otherwise continuous astrophysical background flux, could be a smoking gun signature for the dark matter existence 
in the Universe\cite{rudaz,Ullio,bern,jackson}. 
Recently several groups\cite{bringmann,weniger,tempel} have noted a similar type of spectral line at $130GeV$ of
 energy in publicly available data from the Fermi Large Area Telescope(Fermi-LAT)\cite{fermilat}. 
They consider the annihilation process of dark matter and find the corresponding cross-section,
 $<\sigma v>\approx 2.27\pm0.57^{+0.32}_{-0.51}\times10^{-27} cm^3 s^{-1}( 1.27 \pm0.32^{+0.18}_{-0.28}
\times10^{-27} cm^3 s^{-1})$
  when using the NFW (Einsto) dark matter profile\cite{weniger} in order to expain the 
observed flux. The statistical significance for the line structure is $4.6\sigma$,
or $3.2\sigma$ after the trial factor correction\cite{weniger}. This claim was followed and opposed by many 
groups\cite{mathew,cline,boyarsky,profumo}. Recent update addressing this issue is given in the Ref.\cite{su}.

In this work we would not address the conflicts about the origin of this gamma-ray feature, but 
instead analyze the probability of the Weyl meson to produce this type of spectrum.
The detailed phenomenological implications of the 
Weyl meson has already been discussed in the Ref.\cite{GK}. 
Here we first calculate the annihilation cross-section of the Weyl meson to photon pair and find out the 
coupling parameter for which the observed flux can be explained. Next we calculate the gamma ray flux from 
the decay of the Weyl meson and compare it with the background gamma-ray 
flux. 

This paper is organised as following. In Sec. 2 we summarize the generalized local scale invariant Standard Model. In Sec. 3 we consider the Weyl meson annihilation as
 the source of $130GeV$ gamma-ray line and find the corresponding value of coupling parameter $\lambda_H$ and $\lambda_S$. In Sec. 4 we calculate the 
gamma-ray flux from the decay of Weyl meson and determine the range of parameters within the observational limit. 
Finally we conclude and discuss our result in Sec. 5.
 \section{Generalized local scale invariant theory with one scalar field revisited}
We start with the generalized action for local scale invariant Standard Model (SM) given in Ref.\cite{quantumweyl,shapo}
\begin{equation}
\begin{split}
\mathcal{S}=\int& \sqrt{-\bar{g}} \left(\left[\frac{\beta }{8} \chi ^2+\frac{\beta _1}{4} 
{\mathcal{H}^\dagger} \mathcal{H} \right] {\bar{R}}'\right.
 \left.+{\bar{g}}^{\mu \nu} \left(D_\mu \mathcal{H} \right)^\dagger \left(D_\nu \mathcal{H} \right) +\frac{1}{2} {\bar{g}}^{\mu \nu}\left(D_\mu \chi 
\right) \left(D_\nu \chi \right) \right.\\ 
&\left. -\frac{1}{4}\lambda \chi ^4-\frac{1}{4}\lambda_1 \left[2{\mathcal{H}^\dagger}\mathcal{H}-{\lambda_2} {\chi}^2 \right]^2 \right)
\end{split}
\end{equation}
The couplings $\beta$ and $\lambda_1$ are chosen of the order of unity and the couplings $\lambda,\lambda_2 \ll 1$,
such that $\lambda\sim H_0^2/M_{\rm PL}^2$ and $\lambda_2\sim v^2/M_{\rm PL}^2$. 

Here $\mathcal{H}$ is the Higgs doublet denoted as $\mathcal{H}=\mathcal{H}_0+\mathcal{\hat{H}} $ and $ \mathcal{H}_0,\,\hat{\mathcal{H}}$ are given as
\begin{align}
 \mathcal{H}_0=\frac{1}{\sqrt{2}}\left(
\begin{array}{c}
 0 \\
 v
\end{array}
\right) ,\,
 \mathcal{\hat{H}}=\frac{1}{\sqrt{2}}\left(
\begin{array}{c}
 0 \\
 \phi_3
\end{array}
\right) .
\end{align}
Classically to obtain the gravitational constant we must have
\begin{equation}
\left( \beta \chi_0^2  +\beta_1  v^2\right) \approx M_{pl}^2.
\end{equation}
The scale-covariant curvature scalar ${\bar{R}}'$ is defined in Ref.\cite{cheng,donoghue} and $D_\mu$ is the scale-covariant derivative, 
$ D_\mu=\partial_\mu-fS_\mu$, where $f$ is the gauge coupling constant and $S_\mu $ is the Weyl meson field. By making the quantum expansion 
of the fields around its classical values, 
\begin{equation}
 \begin{split}
  \chi =\chi_{0}+\hat{\chi}, \; \; \;\;\;\;
\phi_{3} =\phi_{3,0}+\hat{\phi_{3}}, \; \; \;\;\;\;
\bar{g}_{\mu\nu}&=g_{\mu\nu}+h'_{\mu\nu} \ ,
 \end{split}
\end{equation}
we can extract the quadratic mass and mixing terms of the fields.
The classical value of the field $\chi$ is taken of the order 
of Planck's mass and the graviton field is redefined, 
\begin{align}
 h'^\beta_\alpha={4 \over \sqrt{\beta \chi_0^2+\beta_1 v^2}} h^\beta_\alpha \ ,
\end{align}
so that its kinetic
energy term gets properly normalized.
Detailed quantum treatment of this model has been discussed
in Ref.\cite{quantumweyl}. 
There it was shown that after choosing a proper gauge one can eliminate the 
mixing terms and identify the 
mass term of Weyl meson as,
\begin{equation} \label{wmass}
M_S^2=f^2\left[\chi_0^2\left(1+\frac{3\beta}{2}\right)+v^2\left(1+\frac{3\beta_1}{2}\right)\right].
\end{equation}
Remaining terms of the action can be arranged in a matrix form as
 $\Phi^T M^2\Phi/2 $, where
\begin{align}
 \Phi=\left(
\begin{array}{c}
 \hat{\chi}  \\
 \phi _3 \\
 h
\end{array}
\right),
\end{align}
is a scalar field triplet and $M$ is the mass matrix, which may be decomposed into unperturbed and perturbed parts as,
\begin{align}
 M^2=M_0^2 +\Delta M^2 .
\end{align}
 The diagonalization of unperturbed 
 mass matrix, considering only leading order terms, gives three eigenvalues which can be identified with the mass of Goldstone type mode, Higgs particle
and graviton, with eigen functions as $\tilde{\chi}$, $\tilde{\phi_3}$ and $\tilde{g}$, respectively. We can write the action in term 
of these particles by inverse transformation. The Weyl-Higgs coupling arises from the kinetic energy term of Higgs boson  and coupling of Higgs with 
gravity in the action. So we can write the interaction Lagrangian as \cite{quantumweyl},
\begin{equation}
\begin{split} 
\mathcal{L}_{int} =\lambda_H \tilde{\phi} _3{}^2S^{\mu }_{;\mu }+\lambda_S{} v \tilde{\phi} _3S^{\mu }
S_{\mu }+ {\lambda_{S}\over 2} \tilde{\phi} _3{}^2S^\mu S_\mu + ...,     
\end{split}    
\label{intlag}  
\end{equation}
where we have defined the couplings as,
\begin{equation}
 \lambda_H={3 \over 4} f \beta_1, \qquad
 \lambda_S={3 \over 2}f^2 \beta_1
\end{equation}
A similar type of interaction between the dark matter particle and Higgs boson arises in the extension of 
SM having extra $U(1)$ gauge symmetry, known as Higgs-portal\cite{baek,choi}.
 In this model $U(1)_X$ gauge boson is the dark matter particle, which interacts with the SM particles through mixing between the complex singlet scalar field and the 
 SM Higgs doublet. 
 
In Ref.\cite{GK} the parameter $f$ was assumed to be sufficiently small and $\beta_1$ sufficiently large
 such that $\lambda_H$ is of order unity. In this case the 
mass of Weyl meson is small and it may produce 
observable effects at colliders such as LHC. The parameter $f$ is directly related to the mass of the 
Weyl meson through the equation(\ref{wmass}). The colliders and the cosmological limit of these parameters has been discussed in the Ref.\cite{GK}.
\section{Weyl meson annihilation to photon pair}
\begin{figure}[!ht]
\begin{center}
\scalebox{0.6}{\includegraphics*[angle=0,width=\textwidth,clip]{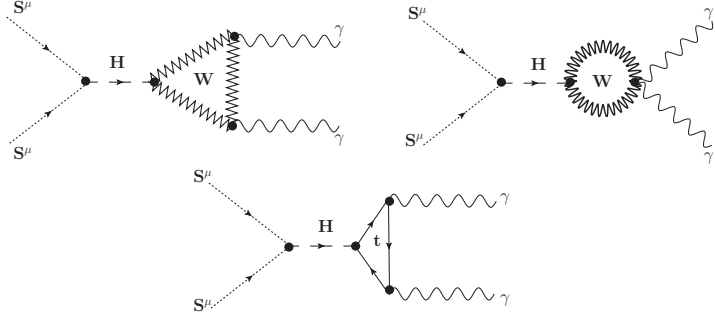}} 
\caption{The Feynman diagram for annihilation of the Weyl meson to produce monochromatic gamma ray line. }
\label{annihilation}
\end{center}
\end{figure}

In generalized locally scale invariant Standard Model, the Weyl meson does 
not interact directly to any of the Standard Model particles except the Higgs boson\cite{quantumweyl,GK}. 
It can ahhihilate to two photons by a Higgs particle through s-channel. Also being massless particles, 
photons do not couple to Higgs boson directly. So the $H\gamma\gamma$ vertex can be generated with 
loops involving $W^\pm$ and charged fermions\cite{higgs}. The dominant Feynman diagrams contributing to this
process are as shown in Fig. \ref{annihilation}. The $SSH$ and $SHH$ vertex are given in Eq.[\ref{intlag}].
\begin{figure}
\centering
\begin{subfigure}{.3\textwidth}
  \centering
  \includegraphics[width=.9\linewidth]{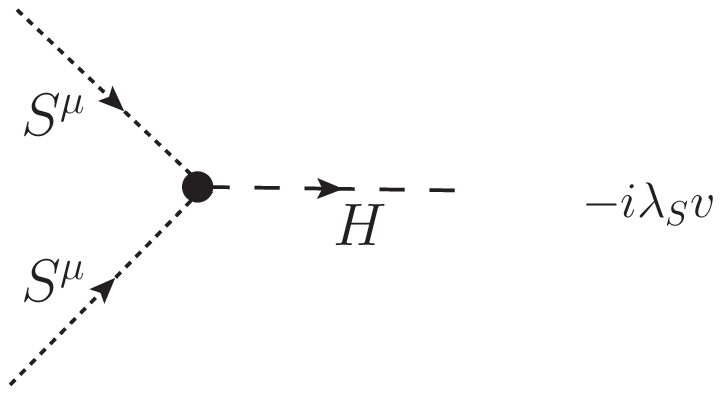}
  \caption{SSH vertex}
  \label{ssh}
\end{subfigure}%
\begin{subfigure}{.3\textwidth}
  \centering
  \includegraphics[width=.9\linewidth]{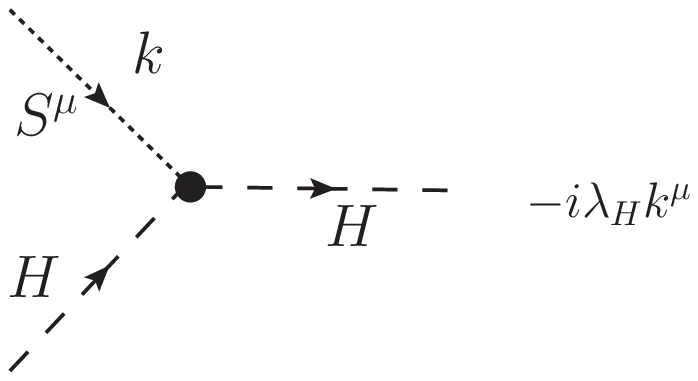}
  \caption{SHH vertex}
  \label{shh}
\end{subfigure}
\end{figure}

Since the final products are massless, we can write the cross section for this process as 
\begin{equation}
<\sigma v>= { 1 \over 16 \pi s}\sum_{spin} |\mathcal M_{SS\rightarrow \gamma \gamma}|^2 
\end{equation}
where $\sqrt{s} $ is the center of mass energy , and
\begin{equation}
\mathcal M _{SS\rightarrow \gamma \gamma}=\lambda_S v \epsilon _\mu \epsilon^\mu 
\left( \frac{i}{s-m_H^2-im_H\Gamma_H}\right) \mathcal M_{H\rightarrow \gamma \gamma}
\end{equation} 
One can write the $\mathcal M_{H\rightarrow \gamma \gamma}$ as follows\cite{higgs}.
\begin{equation}
\begin{split}
\mathcal M_{H\rightarrow \gamma \gamma} &={\alpha_{em} \, s \over 8\pi M_W}[{4\over 3} F_t +F_W]\\
F_t& =-2\tau[1+(1-\tau)f(\tau)],\\
F_W&=2+3\tau+3\tau(2-\tau)f(\tau).\\
 \text{Here}, \, \tau=4M_i^2/s \quad & \text{with} \, i=t,W \, \text{and} \\
f(\tau)&=\left\{ \begin{array}{l c}
\left(sin^{-1}\sqrt{1/ \tau}\right)^2, \quad \tau \geq 1\\
-{1\over 4} \left(ln{1+\sqrt{1-\tau} \over 1-\sqrt{1-\tau}} -i \pi \right)^2 \quad \tau<1
\end{array} \right.
\end{split}
\end{equation}
Therefore, we obtain the cross section as
\begin{equation} \label{sgmav}
\begin{split}
<\sigma v>=&{1\over 16 \pi s} \sum_{spin}|\mathcal M|^2  \\
=&{1 \over 16 \pi s} \lambda_S^2 v^2 \left(2+{1\over M_S^4 }  ( {s\over 2}-M_S ^2)^2\right)
  \frac{|\mathcal M_{H\rightarrow \gamma \gamma}|^2}{(s-m_H^2)^2+m_H^2 \Gamma_H^2}
\end{split}
\end{equation}
where $s=4M_S^2$, is the square of center of mass energy for the non-relativistic case. To produce the 
$130GeV$ gamma-ray line
by two body annihilation process, the mass of each annihilating particle must be approximately $130GeV$.
In Ref.\cite{weniger}, Weigner determined that $<\sigma v>$ should be approximately 
$1.27 \times 10^{-27} cm^3 s^{-1}$ in order to explain the observed Fermi gamma-ray flux. 
Using this value of $<\sigma v>$ and $M_S=130 GeV$ we find from Eq. \ref{sgmav} that the value of coupling  
parameters are , $\lambda_S\approx 2.4$ and $\lambda_H=\lambda_S/2f\approx 1/f$. The parameter $f$, which is 
related to the mass of the Weyl meson through Eq.(\ref{wmass}), is assumed to be very small, such that the Weyl meson 
is not a very massive particle. So we obtain a very large value of SHH vetex, $\lambda_H>>1$, from the annihilation of the Weyl meson.

When we consider the three Higgs production process at LHC, the Higgs-Weyl coupling is 
described by SHH type vertex\cite{quantumweyl,GK}. So the large value of 
$\lambda_H$ obtained from the annihilation leads to the large coupling constant, which will
make the theory perturbatively unreliable. The appearance of such a large coupling disfavour the small mass 
range of the Weyl meson, although it is not entirely ruled out. Qualitatively it will predict a 
relatively large cross-section for three Higgs boson production at LHC. In this paper we do not pursue this 
possibility further. 
\section{Gamma ray flux from the Weyl meson decay}
\begin{figure}[!ht]
\begin{center}
\scalebox{0.4}{\includegraphics*[angle=0,width=\textwidth,clip]{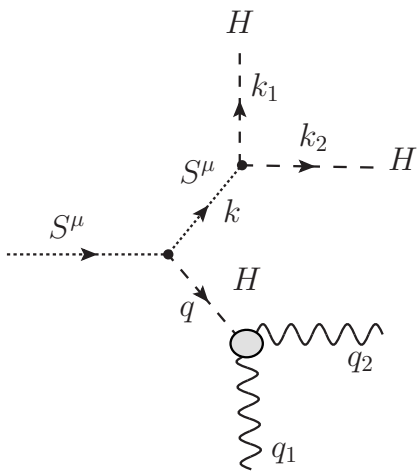}}
\caption{Tree level diagram of the Weyl meson decaying to photon pair. }
\label{wdecay}
\end{center}
\end{figure}

Besides the annihilation, the Weyl meson can also decay into a photon pair. We may observe the gamma ray flux
from the decay of these heavy particles. We consider the tree level decay process of the Weyl meson, shown in Fig.\ref{wdecay}, where blob represent the 
 $W^\pm$ and top quark loop.
The gamma rays are produced 
through the Higgs channel. The Higgs boson will emit 
two photons, each of energy $m_H/2 \approx63$GeV, in its rest frame. The higher energy photons can only be
produced in the boosted frame of the Higgs boson. But the boosting would make the photon spectrum 
broad instead of a sharp peak at $130$GeV.
So the observed peak at $130$GeV would not be explained by this process. However, a lower energy peak at $63 GeV$ is expected.

Here we consider the Higgs boson produced with low kinetic energy such that it emits 
the monochromatic gamma rays of energy $63$GeV. We calculate the flux of 
these $63$GeV photons and compare it with the background gamma rays flux in order to impose a limit on the coupling parameter $\lambda_H$.

The differential flux of photons from the dark matter decay in the galactic core, 
observed from a given direction in the sky, is given as \cite{marco},
\begin{equation}
\frac{{d \Phi}_{\rm dec}}{d E_{\gamma} d \Omega} = \frac{\Gamma}{4\pi} r_\odot \frac{\rho_\odot}{M_S} \int_{\rm{l.o.s}} ds 
\frac{1}{r_\odot} \left( \frac{\rho_{\rm dm}(r)}{\rho_\odot} \right) \frac{d N_{dec}}{d E_{\gamma}}
\label{fluxdec}
\end{equation}
where the coordinate $r$, centered on the Galactic Center, reads $r(s,\psi)=(r_\odot^2+s^2-2\,r_\odot\,s\cos\psi)^{1/2}$,
$r_\odot = 8.5 $ kpc is the distance from the galactic center(GC) to the Sun 
and $\psi$ is the angle between the direction of observation in the sky and the GC. In terms of the galactic latitude $b$ 
and longitude $l$, one has
$
\cos\psi=\cos b\cos l\,.
$
The integral $\int_{l.o.s} ds $ is carried along the line of sight. The DM halo density profile is denoted by $\rho_{dm}(r)$, 
$\rho_\odot\approx 0.4 GeV cm^{-3}$ is the local DM halo density. 

\begin{figure}[!ht]
\begin{center}
\scalebox{0.6}{\includegraphics*[angle=0,width=\textwidth,clip]{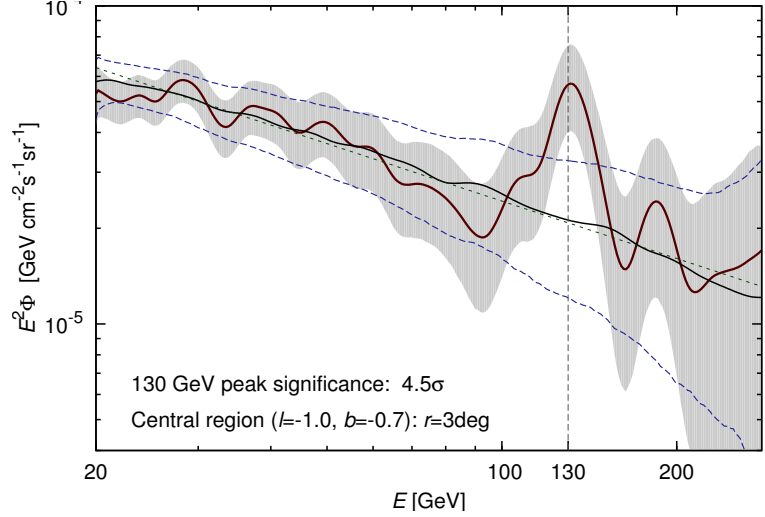}}
\caption{High energy gamma-ray flux together with $95\%$ CL error band as function of photon energy (figure taken from Ref.\cite{tempel}).}
\label{bgrnd}
\end{center}
\end{figure}

We consider the decay of the Weyl meson at rest and $\tau = 1/\Gamma $ is the life time of the Weyl meson.
In the case of $H\rightarrow \gamma \gamma$, the emitted differential photon energy spectrum is 
given as ${dN_{dec} \over dE_{\gamma}}=2 \delta (E-m_H/2)$. We replace the Dirac delta function $\delta(E-m_H/2)$ 
with the Lorentzian Function, such that the gamma rays flux at the peak is given by
\begin{equation}
\frac{{d \Phi}_{\rm dec}}{d E_{\gamma} d \Omega}\lvert_{ E_{\gamma}=63 GeV} = \frac{\Gamma}{4\pi} r_\odot 
\frac{\rho_\odot}{M_S} \int_{\rm{l.o.s}} ds 
\frac{1}{r_\odot} \left( \frac{\rho_{\rm dm}(r)}{\rho_\odot} \right) \frac{2}{\Gamma_H \pi}.
\label{peak}
\end{equation}
The observed gamma-ray flux and background flux varies with energy as shown in the Fig.[\ref{bgrnd}]. 
As the observation does not show any significant peak around $63$GeV, so either the flux of these photons is very small 
compared to the background or the peak lies within the $2\sigma$ error band of background flux. From Eq.(\ref{peak}), this observation
implies that the decay rate of the Weyl meson follows the constraint
\begin{equation}
 \Gamma \leq 4.6\left({M_S\over 100GeV}\right) \times 10^{-27} sec^{-1}\qquad \text{or} \qquad \tau \gtrsim 10^{26}sec.
\label{liftime}
\end{equation}

 This shows that the Weyl meson is a very long lived particle. In our calculations we have used the Einasto DM density profile, 
given as \cite{weniger,navarro,lidia}
\begin{equation}
\rho_{dm}(r) \propto exp\left(\frac{-2}{\alpha_E} \frac{r^{\alpha_E}}{r_s^{\alpha_E}}\right),
\end{equation}
with $\alpha_E=0.17$ and $r_s=20$ kpc.
\subsection{Calculation of the decay rate of the Weyl meson }
In previous section we have obtained the limit on the decay rate of the Weyl meson 
from the observed and background gamma-ray flux. The decay process of the Weyl meson we are considering here is shown in Fig.\ref{wdecay}. 
The amplitude for this process can be written as,
\begin{equation}
 \begin{split}
  \mathcal M&=\frac{\lambda_H\, \lambda_S \,v} {(S_{12}-M_S^2)(S_{34}-m_H^2-i m_H\Gamma_H)} 
\epsilon^\mu \left(g_{\mu\nu}-\frac{k_\mu k_\nu} {M_S^2}\right)k^\nu \mathcal M_{H\rightarrow \gamma \gamma}\\
\sum_{spin} |\mathcal M|^2 &=\left|{2\, \lambda_H^2 \,f\, v \over { M_S^2(S_{34}-m_H^2-i m_H\Gamma_H)}}\right|^2
\left(-S_{12}+\frac{(M_S^2+S_{12}-S_{34})^2}{4M_S^2}\right)\left|\mathcal M_{H\rightarrow \gamma \gamma}\right|^2.
\label{amp}
 \end{split} 
\end{equation}
The decay rate of particle of mass $M_S$ is given by,
\begin{equation}
 \Gamma_S =\frac{1}{2 M_S} \int \sum_{spin}|\mathcal M|^2 d\Phi
\label{decayrate}
\end{equation}
where $d\Phi$ is the phase space integral and for four particle decay process it can be written as \cite{hitoshi,phase}
\begin{equation}
 \int d\Phi= \int \int \frac{d S_{12}}{2 \pi}\frac{d S_{34}}{2 \pi} \frac{\beta_{12}}{8 \pi}\frac{\beta}{8 \pi} \frac{1}{8 \pi}
\end{equation}
with $S_{12}=k^2$, $S_{34}=q^2$ and 
\begin{equation}
 \begin{split}
  \beta_{12}=\sqrt{1-\frac{4 m_H^2}{S_{12}}}, \qquad 
 \beta=\sqrt{1-\frac{2(S_{12}+S_{34})}{M_S^2}+\frac{(S_{12}-S_{34})^2}{M_S^4}}.
 \end{split}
\end{equation}

\begin{figure}[h!]
\begin{center}
\scalebox{0.6}{\includegraphics*[angle=0,width=\textwidth,clip]{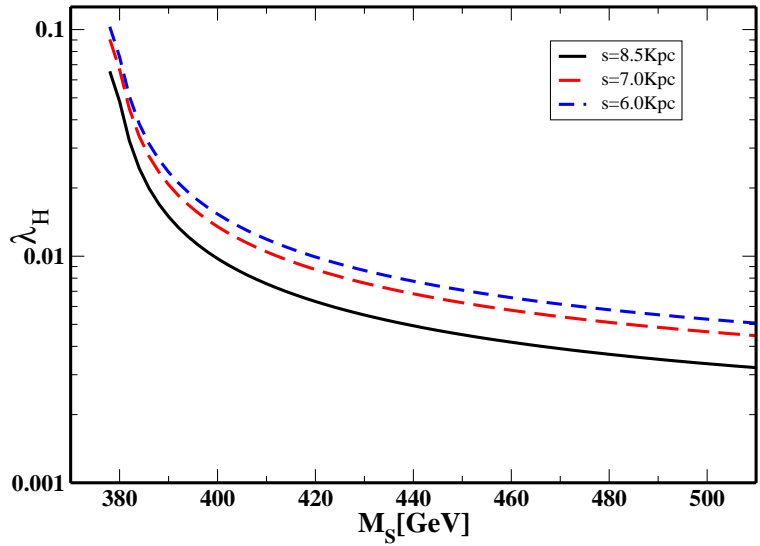}} 
\caption{The maximum value of $\lambda_H$ as the function of the Weyl meson mass for different \emph{l.o.s} distance, $s$.}
\label{dmline}
\end{center}
\end{figure}

Following the constraint Eq.(\ref{liftime}) for the decay rate, $\Gamma$, of the Weyl meson, 
we obtain the limit on the coupling parameter $\lambda_H$ as a function of mass using Eq.[\ref{amp},\ref{decayrate}]. Here we have used $f\approx M_S/m_{pl}$.  
The resulting parameters are plotted in Fig.\ref{dmline}.

In Fig.\ref{dmline}, we give result for different values of the line of sight distance, $s$. The lower mass limit of Weyl meson 
is due the the fact that here we are considering the decay of the Weyl meson into two real Higgs boson
and two gamma-rays such that $M_S \ge 3m_H$. It is clear from the figure that the 
coupling parameter is very small for all mass range of the weyl meson. For small coupling parameter it may be possible that 
the Weyl meson may have decoupled from the cosmic soup immediately after the inflation ended or it may not be
in equilibrium at all, as discussed in Ref.\cite{GK}. So we can not
put any constraint on the relic density of the Weyl meson present in the Universe today. But in colliders the small coupling may produce observable 
effects and we can have the signature of the Weyl meson presence through  the three Higgs production \cite{GK}.

\section{Conclusions and Discussions}
In this work we have studied the implications of the Weyl meson for the gamma-ray line feature 
recently observed by the Fermi-LAT. The Weyl meson is a naturally arising dark matter particle in generalized 
locally scale invariant Standard Model. The cosmological and colliders implications of the Weyl meson has been discussed earlier. 
In this work we have studied the range of coupling parameter and mass of the Weyl meson for the recent Fermi-LAT observation
of gamma-rays coming from the galactic center. This observation shows a peak at $130$GeV which is accounted for by the dark matter annihilation or decay 
at the center of galaxy. We have found that the annihilation of the Weyl meson of mass $130$GeV to produce the required gamma-ray flux need a
very large value of coupling parameter $\lambda_H$. This make perturbation theory unreliable for small mass range of the Weyl meson
in application to collider physics. 
We also calculated the flux of gamma-ray from the decay of the Weyl meson. 
The Weyl meson emits two photons through the Higgs channel, so we expect a peak at $63$GeV of energy. To emit the photons 
of higher energy the Higgs boson need to be boosted, which will make the spectrum broad instead of a sharp peak. The Fermi-LAT 
observations do not show any 
peak around the $63$GeV of energy, which implies that the flux of these gamma rays are small 
compared to the background gamma-ray flux. Hence we can impose a limit on the parameters using the background gamma-ray flux. We have shown that if the 
lifetime of the Weyl meson is greater than $10^{26} sec$, then the flux of gamma-ray produced is comparable to the background flux and 
hence the peak may not be seen by the Fermi-LAT. This constraint implies that for all $M_S>386GeV$, the $\lambda_H$ parameter is always 
less than $0.1$.
\section*{Acknowledgments} 
I thank Prof. Pankaj Jain for many useful discussions and valuable comments. I also thank Mr. Prabhakar Tiwari for discussions. 
I sincerely acknowledge CSIR, New Delhi for financial assistance in the form of SRF
 during this work.
\section*{References}

\end{document}